# Distillation and stripping pilot plants for the JUNO neutrino detector: design, operations and reliability

P. Lombardi[a,b], M. Montuschi[c,d], A. Formozov[a,b,e,f], A. Brigatti[a,b], S. Parmeggiano[a,b], R. Pompilio[a,b], W. Depnering[g], S. Franke[h], R. Gaigher[i], J. Joutsenvaara[j], A. Mengucci[k], E. Meroni[a,b], , H. Steiger[h], F. Mantovani[c,d], G. Ranucci[a,b], G. Andronico[l], V. Antonelli[a,b], M. Baldoncini[c,d], M. Bellato[m], E. Bernieri[n,o], R. Brugnera[m,p], A. Budano[n,o], M. Buscemi[l,r], S. Bussino[n,o], R. Caruso[l,r], D. Chiesa[i,q], C. Clementi[s,t], D. Corti[m], F. Dal Corso[m], X. F. Ding[a,u], S. Dusini[m], A. Fabbri[n,o], G. Fiorentini[c,d], R. Ford[a,w], G. Galet[p], A. Garfagnini[m,p], M. Giammarchi[a,b], A. Giaz[m,p], M. Grassi[a,x], A. Insolia[l,r], R. Isocrate[m], I. Lippi[m], Y. Malyshkin[o], S. M. Mari[n,o], F. Marini[m,p], C. Martellini[n,o], A. Martini[k], M. Mezzetto[m], L. Miramonti[a,b], S. Monforte[l], P. Montini[n,o], M. Nastasi[i,q], F. Ortica[s,t], A. Paoloni[k], D. Pedretti[y,z], N. Pelliccia[s,t], E. Previtali[i,q], A. C. Re[a,b], B. Ricci[c,d], D. Riondino[n,o], A. Romani[s,t], P. Saggese[a,b], G. Salamanna[n,o], F. H. Sawy[m,p], G. Settanta[n,o], M. Sisti[i,q], C. Sirignano[m,p], L. Stanco[m], V. Strati[c,d], G. Verde[l], L. Votano[k].

[a]*INFN — Sezione di Milano, Via Celoria 16, I-20133 Milano, Italy*
[b]*Dipartimento di Fisica, Università di Milano, Via Celoria 16, I-20133 Milano, Italy*
[c]*Dipartimento di Fisica e Scienze della Terra, Università di Ferrara, Via Saragat 1, I-44122 Ferrara, Italy*
[d]*INFN — Sezione di Ferrara, Via Saragat 1, I-44122 Ferrara, Italy*
[e]*Joint Institute for Nuclear Research, 141980 Dubna, Russia*
[f]*Lomonosov Moscow State University Skobeltsyn Institute of Nuclear Physics, 119234 Moscow, Russia*
[g] *Institut für Physik and Excellence Cluster PRISMA, Johannes Gutenberg-Universität Mainz, 55128 Mainz, Germany.*
[h]*Physik-Department, Technische Universität München James-Franck-Str. 1, 85748 Garching, Germany*
[i]*INFN — Sezione di Milano Bicocca, P.zza della Scienza 3, I-20126 Milano, Italy*
[j]*Kerttu Saalasti Institute, University of Oulu, FIN-90014 Oulu, Finland*
[k]*INFN — Laboratori Nazionali di Frascati, Via Fermi 40, I-00044 Frascati (RM), Italy*
[l]*INFN — Sezione di Catania, Via Santa Sofia 64, I-95123 Catania, Italy*
[m]*INFN — Sezione di Padova, Via Marzolo 8, I-35131 Padova, Italy*
[n]*Dipartimento di Matematica e Fisica, Università di Roma Tre, Via della Vasca Navale 84, I-00146 Roma, Italy*
[o]*INFN — Sezione di Roma Tre, Via della Vasca Navale 84, I-00146 Roma, Italy*
[p]*Dipartimento di Fisica e Astronomia, Università di Padova, Via Marzolo 8, I-35131 Padova, Italy*
[q]*Dipartimento di Fisica, Università di Milano Bicocca, P.zza della Scienza 3, I-20126 Milano, Italy*
[r]*Dipartimento di Fisica e Astronomia, Università di Catania, Via Santa Sofia 64, I-95123 Catania, Italy*
[s]*Dipartimento di Chimica, Biologia e Biotecnologia, Università di Perugia, via Elce di Sotto 8, I-06123 Perugia, Italy*
[t]*INFN — Sezione di Perugia, Via Pascoli, I-06123 Perugia, Italy*

[u]Gran Sasso Science Institute, Via Crispi 7, I-67100 L'Aquila, Italy
[w]SNOLAB, Lively, ON, P3Y 1N2 Canada
[x]APC Laboratory— IN2P3, Paris, France
[y]INFN — Laboratori Nazionali di Legnaro, Viale dell'Università 2, I-35020 Legnaro (PD), Italy
[z]Dipartimento di Ingegneria dell'Informazione, Università di Padova, Via Gradenigo 6/b, 35131 Padova Italy

**ABSTRACT**

This paper describes the design, construction principles and operations of the distillation and stripping pilot plants tested at the Daya Bay Neutrino Laboratory, with the perspective to adapt this processes, system cleanliness and leak-tightness to the final full scale plants that will be used for the purification of the liquid scintillator used in the JUNO neutrino detector. The main goal of these plants is to remove radio impurities from the liquid scintillator while increasing its optical attenuation length. Purification of liquid scintillator will be performed with a system combining alumina oxide, distillation, water extraction and steam (or N2 gas) stripping. Such a combined system will aim at obtaining a total attenuation length greater than 20 m @430 nm, and a bulk radiopurity for $^{238}$U and $^{232}$Th in the $10^{-15} \div 10^{-17}$ g/g range. The pilot plants commissioning and operation have also provided valuable information on the degree of reliability of their main components, which will be particularly useful for the design of the final full scale purification equipment for the JUNO liquid scintillator. This paper describe two of the five pilot plants since the Alumina Column, Fluor mixing and the Water Extraction plants are in charge of the Chinese part of the collaboration.

*Keywords: LAB, radiopurity, liquid scintillator, attenuation length, scintillator transparency, light yield, nitrogen purging, large-scale experiments*

## 1 Scientific Motivations

The extraordinary scientific results of the Borexino [1], Daya Bay [2], Double Chooz [3], KamLAND [4] and RENO [5] experiments pave the way for a new generation of multi-kiloton detectors that adopt the Liquid Scintillator (LS) detection technique (JUNO [6], RENO50[7], SNO+ [8], ANDES [9], JINPING[10]).

The Jiangmen Underground Neutrino Observatory (JUNO) is a multi-purpose neutrino experiment, proposed mainly for neutrino mass ordering determination (mass hierarchy) by detecting reactor anti-neutrinos from two sets of nuclear power plants at a 53 km distance. JUNO, deployed in an underground laboratory (700 m overburden), consists in a central detector, a water Cherenkov detector and a top muon tracker. The central detector will be filled with 20 kton of LS and it will be submerged in a water pool, acting as a shield from the natural radioactivity of the surrounding rock. The water pool, in turn, will be instrumented with photomultipliers to act as a Cerenkov detector vetoing cosmic rays background. On top of the water pool, a muon tracker system will accurately measure incoming muons.

The JUNO Liquid Scintillator is a specific organic compound containing molecules featuring benzene rings that can be excited by ionizing particles; it is designed to be composed by Linear Alkyl Benzene (LAB) as solvent, doped with 2,5-Diphenyloxazole (PPO 2.5 g/l) as primary solute, and 1,4-Bis(2-methylstyryl)benzene (bis-MSB 7 mg/l) as wavelength shifter.

Low-background conditions are crucial for the success of JUNO. From the point of view of the LS, this means that the concentration of radioactive impurities inside the mixture should result in an activity of the same level or below the rate of neutrino events. Radiopurity levels are usually specified by the concentration of $^{232}$Th, $^{238}$U and $^{40}$K in the LS and their typical concentration in the environmental sources are listed in Table 1. The baseline scenario, which will be desirable for the detection of reactor antineutrinos in JUNO, assumes a contamination in the range of $10^{-15}$ g/g of U and Th and $10^{-15}$ g/g of $^{40}$K [11] in the LS. A more stringent regime, in the realm of $10^{-17}$ g/g, would instead be needed to accomplish the JUNO neutrino Astroparticle program.

**Table 1** List of the main radioisotopes solute in the organic liquid scintillators with their sources of contamination and the typical concentration of the impurities in the sources [12,13]. In the last two columns are presented the removal strategies used by the main neutrino experiment to reduce the radioimpurities contained in the LS and the JUNO radiopurity requirements[6,11].

| Radioisotope | Contamination source | Typical value | Removal strategy | JUNO requirement |
|---|---|---|---|---|
| $^{222}$Rn | Air and emanation from material | <100 Bq/m$^3$ | Stripping | - |
| $^{238}$U | Dust suspended in liquid | ~$10^{-6}$ g/g | Distillation and Water Extraction | <$10^{-15}$ g/g |
| $^{232}$Th | Dust suspended in liquid | ~$10^{-5}$ g/g | Distillation and Water Extraction | <$10^{-15}$ g/g |
| $^{40}$K | PPO used as doping material | ~$10^{-6}$ g/g | Distillation and Water Extraction | <$10^{-15}$ g/g |
| $^{39}$Ar, $^{42}$Ar | Air | ~1 Bq/m$^3$ | Stripping | - |
| $^{85}$Kr | Air | ~1 Bq/m$^3$ | Stripping | 1 µBq/m$^3$ |

While members of the natural $^{232}$Th and $^{238}$U decay chains are the most common contaminants, also other sources of radioactive impurities for the LS have to be taken into account.

Radioactive impurities can be divided in two main groups according to the process adopted to remove them from the LS. Heavy impurities, such as $^{238}$U, $^{232}$Th and $^{40}$K, can be discarded through distillation and water extraction, while more volatile impurities, such as $^{222}$Rn, $^{39}$Ar, $^{42}$Ar and $^{85}$Kr, through steam or nitrogen stripping. Table 2 displays the concentrations of LS contaminants obtained, after purification, by the main neutrino experiments. It is important to notice that only Borexino and KamLAND achieved the radiopurity standard needed for JUNO.

The JUNO physics program requires reaching an energy resolution (3% at 1 MeV) never achieved before in any large-mass liquid scintillator neutrino experiment. In order to reach the required light collection, the attenuation length has to be comparable to the diameter of the LS acrylic chamber ( A.L.> 20 m at 430 nm [6]). The 430 nm value has been selected since it is in the wavelength region where the PMTs are more sensitive.

The optical performances of the LS are mainly affected by the solvent production methods, and its method of transportation, but the LS attenuation length [14] is influenced also by the different absorbance and cleanliness of each solute (see Table 3). The raw LAB attenuation length, from high quality industrial production, is about 15 m [15], while it could become less than 10 m in standard industrial quality production. For Daya Base pilot plants test a special LAB produced by SINOPEC Jinling Petrochemical Company was selected. Its typical composition is reported in Table 4.

Moreover, any oxidation of the LAB worsens substantially its optical properties, so it is mandatory to avoid any contact between oxygen and the LAB, by keeping any transportation and storage vessel under a nitrogen blanket while removing any air leaks through the connections.

**Table 2** Purification efficacy for different radioisotope in the main LS neutrino experiment (Daya Bay [16], Borexino[17], KamLAND [18] and Double Chooz [19]) in terms of concentrations of radioactive impurities in the LS or event rate.

| Experiment | Radioisotope | Concentration |
|---|---|---|
| Daya Bay | $^{238}$U | <$10^{-12}$ g/g |
|  | $^{232}$Th | <$10^{-12}$ g/g |
| Borexino | $^{238}$U | $(5.3 \pm 0.5) \cdot 10^{-18}$ g/g |
|  | $^{232}$Th | $(3.8 \pm 0.8) \cdot 10^{-18}$ g/g |
|  | $^{40}$K | < 0.42 cpd/100 ton-LS |
|  | $^{222}$Rn | $(1.72 \pm 0.06)$ cpd/100 ton-LS |
|  | $^{39}$Ar | ~0.4 cpd/100 ton-LS (95% C.L.) |
|  | $^{210}$Bi | $(41.0 \pm 1.5(stat) \pm 2.3(sis))$ cpd/100 ton-LS |
|  | $^{85}$Kr | $(30.4 \pm 5.3(stat) \pm 1.5(sis))$ cpd/100 ton-LS |
| KamLAND | $^{238}$U | $(1.87 \pm 0.10) \cdot 10^{-18}$ g/g |
|  | $^{232}$Th | $(8.24 \pm 0.49) \cdot 10^{-17}$ g/g |
|  | $^{40}$K | $(1.30 \pm 0.11) \cdot 10^{-16}$ g/g |
|  | $^{39}$Ar | <$4.3 \cdot 10^{-21}$ g/g |
|  | $^{210}$Pb | $(2.06 \pm 0.04) \cdot 10^{-20}$ g/g |
|  | $^{85}$Kr | $(6.10 \pm 0.14) \cdot 10^{-20}$ g/g |
| Double Chooz | $^{238}$U | <$10^{-13}$ g/g |
|  | $^{232}$Th | <$10^{-13}$ g/g |

In order to test the purification efficiency of the purification process on a LAB based liquid scintillator, it has been decided to build pilot plants with a maximum flow rate of 100 kg/h that will process the LS needed for the filling of one Daya Bay detector in less than 10 days (23.5 m$^3$). In this paper, we focus on the design and operations done during the commissioning phase of distillation and stripping pilot plants, while $Al_2O_3$ filtering system and Water Extraction plant will not be described here since they are in charge of the Chinese part of the collaboration.

Nevertheless, just for comparison, it is worth to mention that one of the plants designed to remove optical impurities and increase the attenuation length of LAB is the $Al_2O_3$ (alumina oxide) filtering system. Alumina is very effective in removing optical contaminants through absorption mechanism. Optical impurities, in principle, could, be removed also through a distillation process by retaining, in the lower part of the column, the high boiling point compounds (such as dust, metal particle and usually oxides) that can affect the light transmittance of the LAB. The last purification system is the Water Extraction plant that is based on the "Scheibel column" design and is intended to remove radioactive contaminants like $^{238}$U, $^{232}$Th and $^{40}$K [29].

In this paper, it is presented the achieved result, obtained with the distillation pilot plant, in removing with high efficiency the optical contaminants.

The continuous many-months operation, implied by the JUNO detector filling, sets severe constraints on the reliability of the final plants. Motivated by these requirements, in Sec. 3 we discuss a reliability model for the distillation and stripping plant based on the data obtained from the operation of the pilot plants during the commissioning and test phases.

**Table 3** Composition of the solvent and solute of the organic LS of the main neutrino experiments (Daya Bay [15, 16, 20], Borexino[13, 17, 24, 26], KamLAND [4, 18, 21,22], Double Chooz[3, 14, 19] and RENO [5, 7, 23]) together with the attenuation length measured at a wavelength of 430 nm after the purification cycle. The attenuation length given for KamLAND was measured at a wavelength of 436 nm.

| Experiment | Solvent | Solute | Attenuation length (m) |
|---|---|---|---|
| Daya Bay | LAB | 1 g/l Gd<br>3 g/l PPO<br>15 mg/l bis-MSB | 14 ± 4 |
| Borexino | PC | 1.45 g/l PPO | ~10 |
| KamLAND | 80% Dodecane<br>20 % PC | 1.36 g/l PPO | 12.7 ± 0.4 |
| Double Chooz | 80% n-Dodecane<br>20 % o-PXE | 4.5 g/l Gd-(thd)$_3$<br>0.5%wt Oxolane<br>7 g/l PPO<br>20 mg/l bis-MSB | 7.8 ± 0.5 |
| RENO | LAB | 3 g/l PPO<br>30 mg/l bis-MSB<br>1 g/l Gd | >10 |

**Table 4** Composition of special LAB used for the commissioning of the distillation and stripping test at Daya Bay Neutrino Laboratory produced by SINOPEC Jinling Petrochemical Company. LAB is a mixture of compound that can be expressed in terms of n in the form of $(C_6H_5)-C_nH_{2n+1}$.

| Components<br>$C_6H_5C_nH_{2n+1}$ | Concentration<br>% |
|---|---|
| n = 9 | 0 % |
| n = 10 | 10 % |
| n = 11 | 35 % |
| n = 12 | 35 % |
| n = 13 | 20 % |
| n = 14 | 0 % |

## 2 Distillation and stripping pilot plant overview

Distillation and stripping technologies are widely used for purification of Liquid Scintillators in large-scale neutrino experiments. In this respect, the JUNO LS purification system has a particularly difficult task since both excellent radiopurity and extraordinary optical quality have to be reached. In addition, a high production rate must be achieved together with compliance with Chinese and European safety regulations. In the following sections, we describe the main features of the distillation and stripping pilot plants, installed at the Daya Bay site. Pilot plants design, construction and operation has been a crucial step to understand and prove purification efficiency. All the knowledge and feedback acquired in this pilot test phase will be crucial to optimize and further upgrade the design of the full-scale plants of JUNO experiment.

### 2.1 Distillation plant

Distillation plant is used to remove form the raw LAB the heaviest impurities (mainly $^{238}$U, $^{232}$Th and $^{40}$K) and to improve its optical property in terms of absorbance spectrum and attenuation length in the 350 nm – 550 nm wavelength region. This process is based on the heat and mass transfer between a liquid and a gas stream, due to the equilibrium conditions reached on each stage of a distillation column. These conditions depend on the difference of volatility between the constituents of the input stream and on the temperature and pressure in the column. The low volatility components are concentrated in the bottom of the system, while the high volatility ones on the top.

The distillation is carried out with counter-current flow of the liquid and gaseous LAB in a 7 m high and 2000 mm wide column containing 6 sieve trays (see Fig. 1 and Table 5). In particular, the height of the column and trays number affect separation capability, while total flow rates determine the width of the column.

The three principal components of the distillation system are the column, the reboiler and the total condenser. Liquid LAB is fed to the column at a flow rate of about 100 l/h in the middle tray section (1 in Fig. 1), after being preheated (~160 °C) in the vapour condenser (2 in Fig. 1) on the top of the column. The liquid stream, falling down by gravity through the sieve trays, reaches the reboiler, which evaporates the liquid with a 15 kW$_{th}$ electric heater (immersed resistors) generating the counter current flow of vapor. Temperature in the reboiler is around 200 °C depending on the column actual pressure and the LAB chain composition. The trays are designed in order to establish an intimate contact between the liquid stream and the gas stream for a sufficient period of time allowing the heat and mass transfer between the phases. This process enriches the liquid stream in the less volatile components (in particular $^{238}$U and $^{232}$Th and heaviest impurities) and decreases the temperature of the vapors. The liquid and vapor flows must be kept within a limited operating range to assure a good contact surface on the sieve trays.

The top of the distillation column features the total condenser (2 in Fig. 1), cooled by the LAB input flow, where the LAB vapors are liquefied. In this design, the total condenser has the function of energy recovery. The product liquid stream is then split by the condenser itself in two currents, one inserted back inside the column as a reflux flow (to increase the efficiency of the distillation process) and the other directed to the water based heat-exchanger (3 in Fig. 1) for the sub-cooling to ambient temperature and then sent to the product tank.

The distillation pilot plant is operated with a nominal reflux ratio of 25%, adjusted varying the product flow, and a 2% of the input flow discharge from the bottom of the column in order to get a good compromise between the product purity and a reasonable throughput [12].

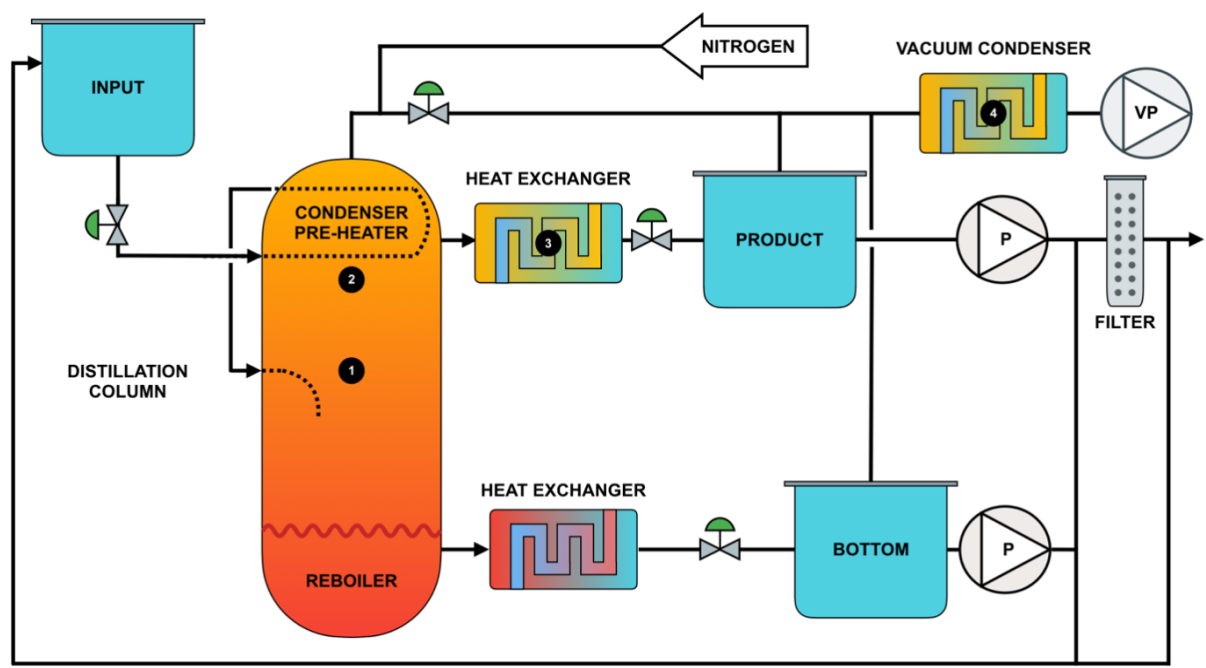

**Fig. 1**. Distillation pilot plant sketch (not in scale). The raw LAB from the input tank falls by gravity through the top of the column where is pre-heated by the LAB vapour inside the total condenser installed right on top of the column (2). It is then, at a temperature of roughly 160 °C, sent to the column at the middle tray (1) where it falls down in the electric reboiler (~200 °C) integrated in the distillation column itself. The reboiler generates heat with submerged electric resistances. The LAB vapours are then condensed in the top of the column and split in the product stream and in the reflux stream (~ 25% of the product stream). The flow of the distilled LAB is then cooled down at ambient temperature (3) and collected in the product tank. The discharge flow (~ 2% of the input stream) from the reboiler and sent to its collecting tank after being cooled down at ambient temperature. The pressure inside the distillation column, the product tank and the bottom tank is kept constant at a value of 5 mbar$_a$ with a scroll vacuum pump (VP) and a continuous purge of nitrogen. The distilled LAB can be then pumped back by a diaphragm pump (P) to the input tank, so to distil it in internal loop mode, or sent to the next purification step passing through a 50 nm pore filter. In order to recover the LAB discharged from the bottom of the column it can be pumped back to the input tank.

The distilled LAB is then sent to the next purification process through a 50 nm pore filter in order to retain any dust or metal particles already present or introduced in the stream by the plant itself.

The entire plant is kept under a $N_2$ blanket provided by a continuous gas flow to avoid any oxidation inside the column, thus also reducing the risk of fire. The incondensable gas stream, if present, is then removed from the top of the column by a dry scroll vacuum pump, in order to keep a constant pressure of 5 mbar inside the column, passing through a vacuum condenser (4 in Fig. 1) to liquefy any possibly LAB vapor dragged by the nitrogen flow.

The plant can be operated in two different ways: the internal loop mode, where the LAB from the product tank and the filter, is sent back to the feed tank, and the continuous mode where the feed tank (1 m$^3$) is constantly filled with raw LAB and the distilled LAB is sent from the product tank (0.5 m3) to the next purification step continuously. The first configuration is used only in the start-up phase of the plants or if a stop of the detector filling occur, while the second is the production mode.

Table 5 Main operational parameters for the different features of the distillation pilot plant tested at Daya Bay.

| Feature | Value |
|---|---|
| Height | 7 m |
| Diameter | 200 mm |
| Number of trays | 6 |
| Pressure | 5 mbar$_a$ |
| Temperature in the reboiler | 200 °C |
| Temperature in the top of the column | 160 °C |
| Input flow | 100 l/h |
| Reflux flow | 25 l/h |
| Discharge flow | 2 l/h |
| Nitrogen flow | 2 kg/h |
| Electrical Power for the heater | 20 kW$_{th}$ |
| Cooling Power | 14 kW$_{th}$ |
| Feed tank Volume | 1 m$^3$ |
| Product tank Volume | 0.5 m$^3$ |
| Bottom Tank Volum | 0.5 m$^3$ |

The solutions listed below are adopted in order to achieve better performances in terms of removal of the radioactive impurities, energy saving and cleanliness.

- Sieve Trays: they have the simplest design among various tray types and feature neither mechanical moving parts nor welding, which permits an easy and effective cleaning. The trays have 55 holes with a diameter of 12 mm to allow a good contact surface between the vapor and the liquid phase and no down-comer in order to avoid any parts that could be difficult to clean. The size and number of the holes in trays are based on nominal flow rates of vapor rising up and liquid falling down the column. If the flows are too high or too low, bypassing occurs, reducing the contact surface and the stage efficiency.
- Total Condenser: the condenser is positioned directly on the top of the column in order to reduce the size of the plant. Moreover, the LAB vapor is cooled down by the LAB liquid input stream. The pre-heating of the LAB input stream permits an energy recovery of the order of 10 kW$_{th}$, while also avoiding the destabilization of the column temperature profile, due to the insertion of cool fluid in the middle.
- Vacuum distillation column: in order to achieve better purification performances, the distillation process pressure is kept below 5 mbar$_a$, increasing the difference between the vapor pressure of the LAB and that of heaviest impurities. A low pressure inside the column reduces the LAB boiling temperature (less than 200 °C), decreasing effectively the risk of thermal degradation of LAB.
- At the design conditions of 100 l/h feed and reflux ratio 1, the six-tray column was predicted to have four theoretical stages based on design correlations.

## 2.2 Stripping plant

After LAB purification through Alumina and Distillation plants, liquid scintillator is prepared by online mixing of purified LAB with the right percent of a Master Solution mixture (MS). MS is a concentrated solution of LAB + 100 g/l PPO and 280 mg/l bisMSB, pre-purified in a dedicated plant (water extraction in batch mode). Liquid scintillator stream is finally processed through Water Extraction and Stripping plants.

The gas stripping is a separation process in which, one or more dissolved gases are removed from the liquid phase and transferred to the gas phase by the desorption mechanism. For example, radioactive gases (mainly $^{85}$Kr, $^{39}$Ar and $^{222}$Rn) and oxygen (which potentially decreases the light yield due to photon quenching) can be removed from the scintillator mixture by stripping it with a variable mixture of superheated steam and nitrogen in counter current mode. The stripping pilot plant was designed to measure the process efficiency with superheated steam, $N_2$ or a combination of the two in order to identify the best configuration for the future full size plants.

In this paragraph

The pre-heated liquid stream (2 in Fig. 2) enters the stripping column (1 in Fig. 2) from the top and falls down by gravity through an unstructured packing (Pall rings) that permits a high contact surface between the liquid and the gas coming from the bottom of the column (Fig.2 and Table 5).

The concentrations of dissolved gases in the two streams ($y_i$ for the liquid phase and $x_i$ for the gas mixture) vary in each stages of the column, depending on the equilibrium conditions between liquid and gaseous flows, as governed by the Henry law:

$$y_i \cdot p_t = H_i \cdot x_i$$

where $p_t$ is the process pressure and $H_i$ the Henry's law constant that depends on temperature, pressure and the composition of the streams at the i-th theoretical stage. In order to keep the pressure gradient constant inside the stripping column, the steam is condensed in vacuum condensers, while the incondensable constituents of the gas stream are discharged by a scroll vacuum pump (3 in Fig. 2).

The Henry constant, in combination with the molar fraction, determines the maximum ratio between liquid flow $L$ and gas flow $G$. By applying the mass balance to the column:

$$\frac{L}{G}\bigg|_{max} = \frac{x_2 - x_1}{y_1 - y_2}$$

The optimal liquid-gas ratio is higher than 70% of the maximum L/G ratio, to avoid large gas flow and high pressure loss inside the column, and lower than 85% of L/G max, not to increase too much the height of the column due to a minor driving force between liquid and gas.

The stripped liquid, collected in the bottom of the column, is sent to the product tank (0.5 m$^3$) by a pump through a water based heat exchanger to lower its temperature, and through a 50 nm filter used to retain the dust and the particulate that can be released by the plant itself.

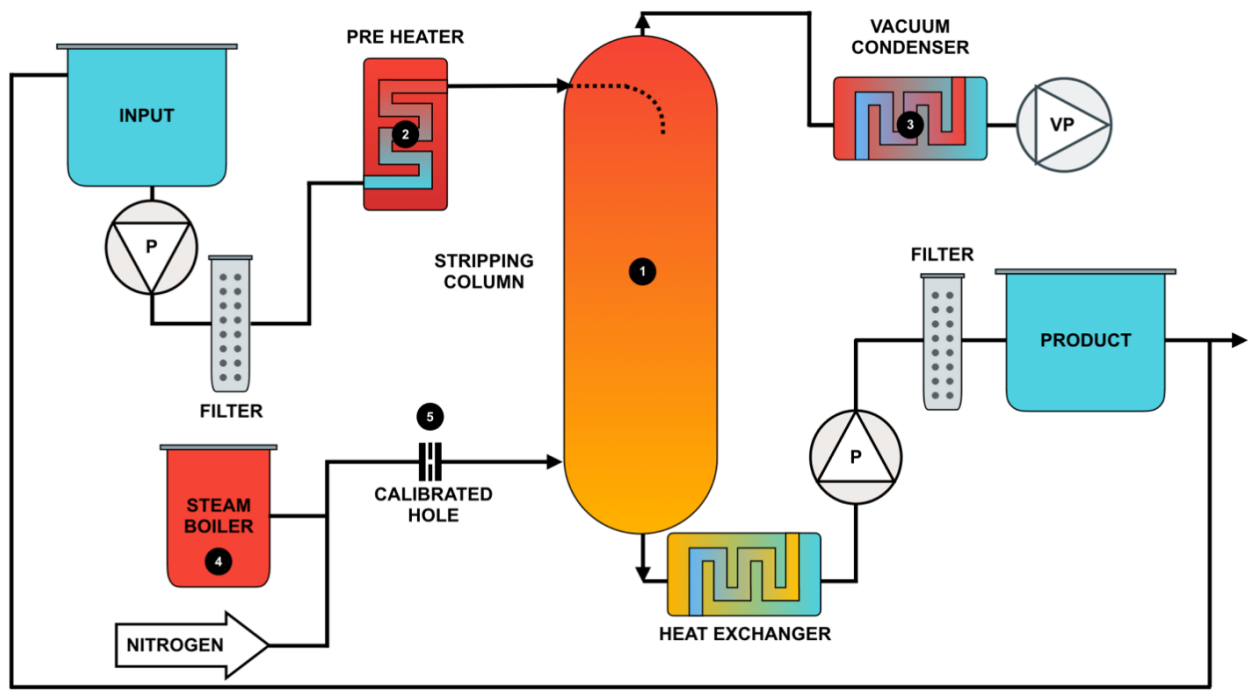

**Fig. 2**. Stripping pilot plant sketch (not in scale). The LAB, collected in the input tank from the previous purification steps, is pumped by a diaphragm pump (P) to the top of the stripping column after being filtered through a 50 nm pore filter and preheated at 80 °C in the oil based heater (1) in order to avoid the condensation of steam inside the liquid stream. The gas flow is an adjustable mix of nitrogen and steam produced inside the electrical steam boiler (2) at a pressure > 150 $mbar_a$ kept constant by the continuous flow of the steam through a calibrated orifice (5) to the stripping column (1). The stripping column is filled with Pall rings in order to maximize the contact surface between the liquid and the gas stream. The stripped LAB is then collected in the bottom of the column and sent to the product tank after being cooled down in a water based heat exchanger and filtered. The liquid can be then sent back to the input tank or pumped out to the filling station of the detector. The gas flow is discharged by a scroll vacuum pump (VP) after being cooled down in the vacuum condenser (3) in order to condense the steam remove the water from the stream.

The nitrogen used is carefully purified with active carbons at cryogenic temperature to reach low concentration of radio-contaminants, because they set a lower limit for the radiopurity that can be achieved by gas stripping.

The steam flow is produced in a 50 l volume steam boiler (4 in Fig. 2), at a temperature around 70 °C (pressure around 300 $mbar_a$) using ultrapure water from the high purity water plant of Daya Bay [16]. Its flow is controlled by a calibrated orifice hole with a diameter of 0.3 mm (5 in Fig. 2) located between the heater and the needle valve installed on the superheated steam line before the column. Possible condensation of steam in the column is avoided by operative solution. The LS, and the entire column as a consequence, is pre-heated at 90 °C. This temperature is 20 °C more than the production temperature of the steam at even higher pressure of the column (300 mbar vs 250 mbar). These precautions bring the steam a superheated steam at soon as it enter the column. The superheated steam could therefore be treated like a gas with no phase separation.

This plant can be operated in the internal loop mode (during the start-up operations and self-cleaning procedures) and in continuous mode where the purified LAB is sent, after stripping, from the product tank (0.5 m3) to the filling station of the Daya Bay detector.

In order to reach the purity and optical standards needed for JUNO, the following design options have been adopted.

- Unstructured Packing: the column is filled with AISI316 Pall rings to increase the contact area between the liquid and gas stream. They have been electro polished and effectively cleaned before the installation inside the column with an ultrasonic bath.
- Stripping under vacuum: the reduced pressure can improve the efficiency per theoretical stage of gas stripping. On the other hand, the inter-facial mass transport rate is substantially reduced in the absence of gas flow. In a stripping column of fixed size, there is an optimal pressure for gas stripping: reducing pressure increases the efficiency per theoretical stage, but also decreases the number of theoretical stages. The optimal pressure for our stripping operations is between 150 and 250 mbar$_a$.
- Steam: the use of steam instead of Nitrogen (the Borexino choice [13]), has two advantages. Firstly, it is generally easier to produce ultrapure water than $N_2$ with a low content of radioactive contaminant reaching a concentration of $^{222}$Rn < 3.4·10$^{-6}$ Bq/kg and a very low concentration in $^{39}$Ar and $^{85}$Kr. [24]. Moreover, using Nitrogen as a stripping gas requires adopting an exhaust system to displace it in a sufficiently well ventilated place. The amount of dissolved water in LAB at 100% saturation at atmospheric pressure and room temperature is ~200 ppm. Stripping at ~250 mbar$_a$ (even if at a temperature around 90 °C) reduce the amount of water dissolved in the LS after the cooling heat exchanger. The measured content of water in LS after steam stripping was ~50 ppm and it does not represent an issue for JUNO experiment.
- LS pre-heater: as already mentioned, in order to avoid any condensation of steam in the LS stream, the LS is heated at a temperature of 90 °C. Increasing the temperature give also the advantage to enhance the stripping efficiency.
- At the design conditions the 4 m, unstructured packed column was predicted to have three theoretical stages.

Table 6 Main operational parameters for the different features of the stripping pilot plant tested at Daya Bay.

| Feature | Value |
| --- | --- |
| Height | 7 m (4 m of unstructured Packing) |
| Diameter | 75 mm |
| Packing Material | AISI 316 Pall rings |
| Pressure | 150 – 250 mbar$_a$ |
| Input LAB Flow temperature | 90 °C |
| Steam temperature | 70 °C |
| Input LAB flow | 100 l/h |
| Steam flow | 100 g/h |
| Nitrogen flow | 1 Nm$^3$/h |
| Electrical Power for the heater | 10 kW$_{th}$ |
| Cooling Power | 5 kW$_{th}$ |
| Feed tank Volume | 0.5 m$^3$ |
| Product tank Volume | 0.5 m$^3$ |

## 2.3 Common Features

In order to avoid any contamination due to the dust, dirt and oxide particles which could be released into the detector or liquid handling systems, it is mandatory to use electro-polished 316L stainless steel and special cleaning process. Following we describe the cleaning procedures adopted to treat all the parts of the distillation and stripping pilot plants such as pipes, tanks, valves, pumps and sensors.

The desired cleanliness standard for the plant is MIL STD 1246 Level 50 [25], which defines limits on the residual particulate size distribution. This goal assumes the scintillator causes particulate wash-off similar to water, and that Class 50 is the acceptable level for the scintillator, assuming the remaining particulate has a radioactivity similar to dust. Hopefully, the second assumption is not true, and the remaining particulate is mostly metallic (i.e. less radioactive than dust), resulting in very conservative specifications for the lines.

The procedure has followed these steps [26]:
- detergent cycle, to remove oil, grease and residuals with Alconox Detergent 8 or equivalent (concentration 3% at 60 °C);
- Ultra-Pure Water (UPW) cycle for rinsing (Until resistivity > 4 MΩ cm)
- pickling and passivation;
- UPW cycle for final rinsing (Until resistivity > 14 MΩ cm.)

Small parts have been cleaned in ultrasonic baths, while bigger parts with appropriate methods, like spray balls or immersion.

Moreover, at the end of each plant we decided to install a (pre-wetted) ultra-filter with the nominal pore diameter of 50 nm, to retain any kind of particles that can be released by the plant itself.

Specific attention is given to avoid leaks through the connections. In particular, all large flanges and the ones withstanding ambient temperature are sealed with Ansiflex gaskets or Viton Teflon coated gaskets, while in the high temperature parts of the plant the tightness is assured by using metal loaded TUF-STEEL gaskets. All process line connections are orbital-welded or TIG-welded using low thorium content electrodes. Where welding is not possible, metal gasket VCR fittings are used. Moreover, all instrument probes are connected to the plant with vacuum tight fittings for high seal, and stainless steel diaphragm sealed valves are used throughout the system. (The overall integral leak rate of each plant was proved to be less than $10^{-8}$ mbar-l/s by means of a He leak detector).

The skids have to meet safety European and Chinese requirements in terms of certification of seismic safety. A Hazop procedure was used to identify potential problems during operations and led to modifications for the sensing and alarming parts of the system. In order to avoid the prescription of the PED directive, rupture disks are installed to assure in every tank a local pressure lower than 0.49 $bar_g$. In particular, rupture disks are designed to be operative between full vacuum up to the trigger point of 0.45 $bar_g$.

All the electric equipment are under ATEX specification [27], in Class 1 Zone 2 T2, to prevent any fire risk since the LAB temperature is above its flash point in the distillation plant.

All the process pumps used are volumetric diaphragm pumps with Teflon membranes, installed in the lower part of the plants in order to help the pump priming and to avoid the cavitation in compliance with instrument NPSH. The pumps used to move liquid from a low-pressure tank to an ambient pressure tank are compressed air driven DEBEM pump, while in all the other cases we use motor driven PROMINENT pump.

These purification plants need a very stable and reliable Distributed Control System (DCS) to adjust the purification parameters and to assure the safety of both the plants and the operators, considering the elevated temperatures that exist in the plant (in distillation mode) and the enclosed environment in which the plants are located. The purification system has to be under the control of a master system that provides, for 24-h/day operation, alarm notification, and automated shutdown in case of problems.

It has been decided to adopt a Siemens system for distributed automation because it guarantees good performances in terms of reliability and a modular and safety oriented design. Moreover, it can be used in hazardous areas (ATEX Zone 2). The CPU module chosen is the 1512SP-1P. It assures different communication options between the PLC and the PC with the possibility to integrate a channel specific diagnostic.

The DCS can be controlled and monitored via a SCADA application, designed integrating an operator friendly User Interface (UI), with the purpose to permit a quick learning of the plant operations and to understand and solve easily the cause of any alarms generated by the DCS. This application runs on a Local PC, where it saves all the processes parameter values every minute. It is linked to the PLC via an Ethernet connection.

The general UI is divided in three tabs: an overview of the plant (see Fig. 3), an alarm panel and a trend panel.

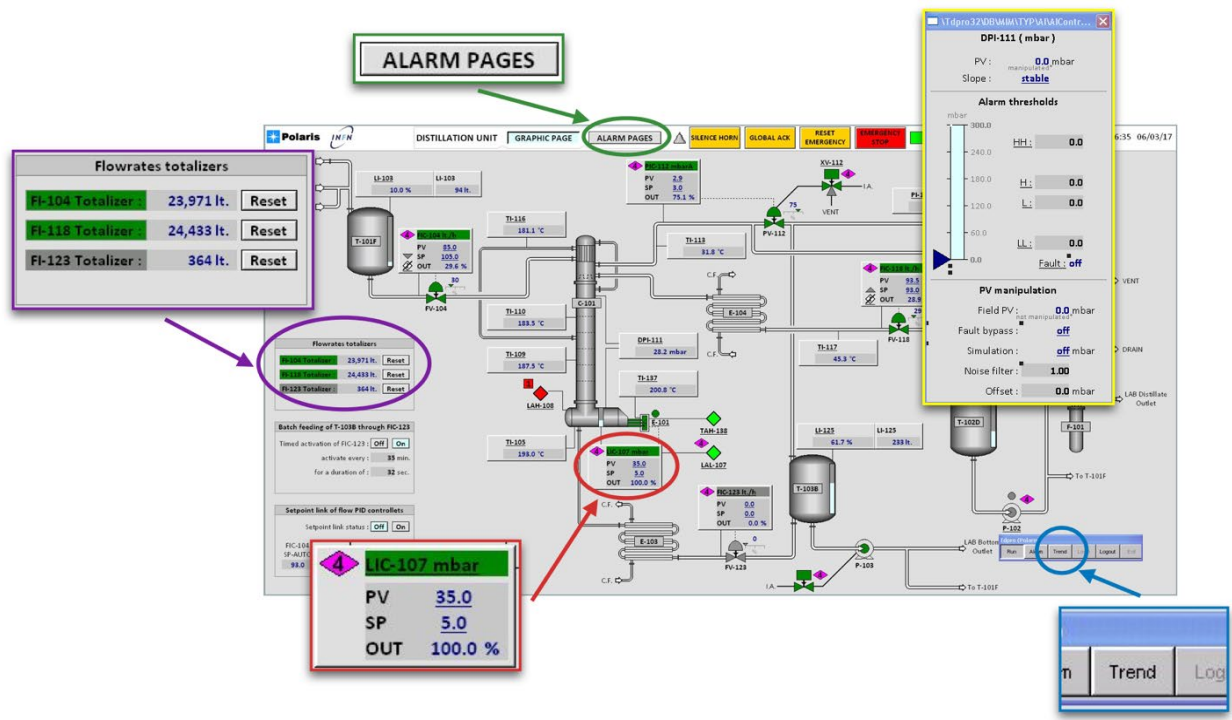

**Fig. 3.** The slow control User Interface (UI) is designed in order to guarantee a fast identification of the values of the process parameter. It is possible to set each instrument alarm thresholds (HighHigh, High, Low and LowLow) and to adjust the process parameters with the instrument panel. In the Alarm Pages tab are collected all the previous and active alarms and it is possible to examine the progress of each instrument value with the trend graph. The slow control User Interface (UI) shows also the flowrates totalizer keeping always under control the amount of processed LS.

In the first tab, the core of the UI, it is possible to set the process parameters and the alarm thresholds, open and close the automatic valves and turn the pumps on and off. Here the measured values of each instrument connected to the DCS are also displayed.

The second panel collects all the alarms that are active or were active, but not acknowledged, while in the last it is possible to monitor the trend over time of the process values, which are also saved on the PC.

The DCS manages also part of the safety rules that prevent any damage to the plant and to the operators. In particular, it prevents the switch-on of the equipment if the proper conditions are not satisfied: for example if the LAB level in the distillation reboiler is not high enough the heaters cannot be turned on.

It is foreseen also an account based system in order to establish a hierarchy between users of the DCS and to give the privileges of change the settings only to expert operators and just monitoring capabilities to the guests.

## 3 Reliability

The JUNO purification plants will have to face the highly demanding challenge of assuring a constant delivery of purified LS for the entire filling period. A further hurdle arises from the fact that the last stages of the purification process will take place in the underground laboratory with the aim of minimizing the length of the pipe from the stripping plant to the filling stations and of reducing the risk of contaminating the purified LS. In this scenario, the replacement of LS in case of failure of the purification process will be almost unfeasible. For these reasons, a reliability assessment is mandatory in order to identify the less resilient components and possibly maximize the robustness and safety of the whole purification system. Essentially It has been decided to use the experience gained by the design and operations done on the pilot systems in order to develop a reliability study of the future JUNO purification plants. In the following the calculations done for pilot plants are given. The collected statistic after 2 years of pilot plants operations is in good agreement with the expectations.

Reliability is generally defined as the probability $R(t)$ of successful performance under specified conditions of time and use and it is related with the failure rate $\lambda(t)$ of every single component of the system [28]:

$$R(t) = e^{-\int \lambda(t)dt} \qquad (1)$$

The lifetime of a component can be divided in three stages: the infant mortality period when the failure rate is not constant and decreases rapidly with time, the life period when the failure rate is considered constant and the wear out period where the failure rate increases rapidly due to ageing of the component.

In our case, the infant mortality period is considered finished after the commissioning of the plants, so we consider the components inside the constant failure rate period and it is possible to use failure rates from literature or from similar plants.

The total reliability of a complex structure can be calculated using the probability theory breaking down the entire system in simpler modules or subsystem arranged in series or in parallel [28].

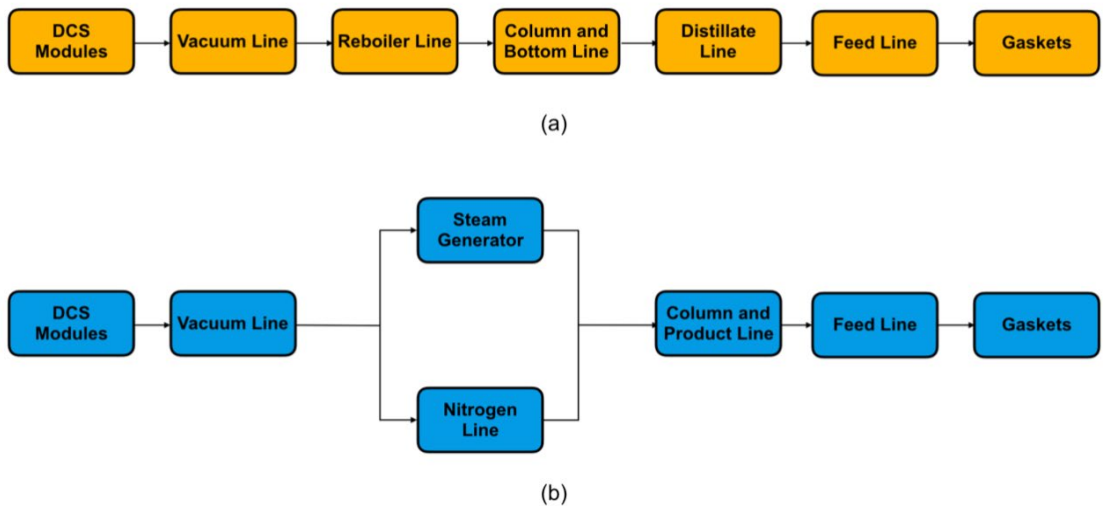

**Fig. 4.** Subsystem of the distillation pilot plant (a) and stripping pilot plant (b). The distillation pilot plant total reliability can be calculated as the product of the reliability of the single subsystem because all the plant works in series one to each other. While the stripping plant reliability can be evaluated as the product of all the other subsystem with the reliability of the subsystem composed by the Steam Generator and the Nitrogen.

In the distillation plant all the subsystems are arranged in series (see Fig. 4a), implying that the total reliability can be estimated using equation (2). In the stripping pilot plant one stage involves a parallel between the Steam Generator and the Nitrogen Line (see Fig. 4b): therefore the total reliability $R_{tot}$ can be evaluated by combining the reliability of the Steam Generator plus Nitrogen Line subsystems in parallel using equation (3) with the reliabilities of the remaining components:

$$R_{tot} = \prod_i R_i \qquad (2)$$
$$R_{tot} = 1 - \prod_i (1 - R_i) \qquad (3)$$

The failure rate of each components, listed in Table 7, are combined with the previous equations to get the final reliability and the Mean Time Between Failure (MTBF) (see Table 8) in order to estimate the number of stops for the plants, considering the reliability of the external utilities, provided by the lab (i.e. chiller, water supply, nitrogen supply). The reliability of the hand-operated valves is set to 1. The MTBF (measured in hours) is correlated with the failure rate through the following equation, when $\lambda(t)$ is considered constant:

$$MTBF = \frac{1}{\lambda}$$

**Table 7** List of the main components of the distillation and stripping pilot plant used and their failure rate given by the production company and from Borexino experience.

| Component | Failure Rate $\lambda$ (fail/$10^6$ h) |
|---|---|
| Pressure sensor | 1.7 |
| Regulating valve | 30 |
| Heat exchanger | 20 |
| Vacuum pump | 15 |
| Level sensor | 12 |
| Thermocouple | 10.1 |
| Level switch | 4.5 |
| On/Off valve | 20 |
| Rupture disk | 13.5 |
| Centrifugal pump | 20 |
| Flow meter | 5 |
| Filter | 1 |
| Gaskets | 0.2 |
| DCS module | 1 |
| Filter | 1 |
| Steam generator | 50 |
| Pressure reducer | 0.3 |

Due to a less complex system and less physical objects inside the plant, the stripping system has a lower failure probability than the distillation plant. Therefore, it has a longer MTBF meaning a longer continuous activity between two stops for maintenance. Finally, considering 6 months of continuous working time to fill the JUNO detector, we will have 2 stops in 6 month of continuous operation for each plant (stripping and distillation) with a mean down time estimated of 36 h/failure, with a total of 3 days of stops for each plant.

**Table 8** Probability of successful performances (R) and Mean Time Before Failure (MTBF) in months calculated for each subsystem composing the distillation and stripping pilot plant and for the entire plants. The model used for the calculation is shown in Fig. 4 and the failure rate for each component of the subsystem are listed in

| | Line description | R | MTBF ($10^3$ h) |
|---|---|---|---|
| Distillation | Vacuum line | 0.637 | 30.9 |
| | Reboiler line | 0.797 | 23.8 |
| | Column + bottom | 0.576 | 7.9 |
| | Distillate line | 0.665 | 7.9 |
| | Feed line | 0.722 | 15.8 |
| | Gaskets (200) | 0.916 | 14.4 |
| | DCS modules | 0.961 | 98.6 |
| | Total | 0.124 | 2.2 |
| Stripping | Vacuum Line | 0.835 | 36.7 |
| | GV | 0.698 | 12.2 |
| | Column + product | 0.524 | 5.8 |
| | Feed line | 0.613 | 8.6 |
| | Nitrogen line | 0.978 | 98.6 |
| | Gaskets (150) | 0.936 | 19.4 |
| | DCS modules | 0.961 | 98.6 |
| | Total | 0.235 | 2.9 |

## 4    From designing to commissioning

In 2014-2015 the design and the construction of the JUNO purification pilot plants was started, with the aim to test them in the Daya Bay Laboratory and to find the optimal process parameters for the design of the final full scale plants.

During the period between 2015-2016, the construction work for the distillation and stripping plants was carried out in conjunction with Polaris Engineering (MB, Italy) under the supervision of the Istituto Nazionale di Fisica Nucleare (INFN) crew.

The plants were designed and built as a skid-mounted system (see Fig. 5) for transportation flexibility in China (they fit into two 2.15m x 2.4m x 7m skids). INFN reviewed and approved all materials, equipment selections and fabrication methods to ensure that the system was leak tight and had the possibility to be completely cleaned.

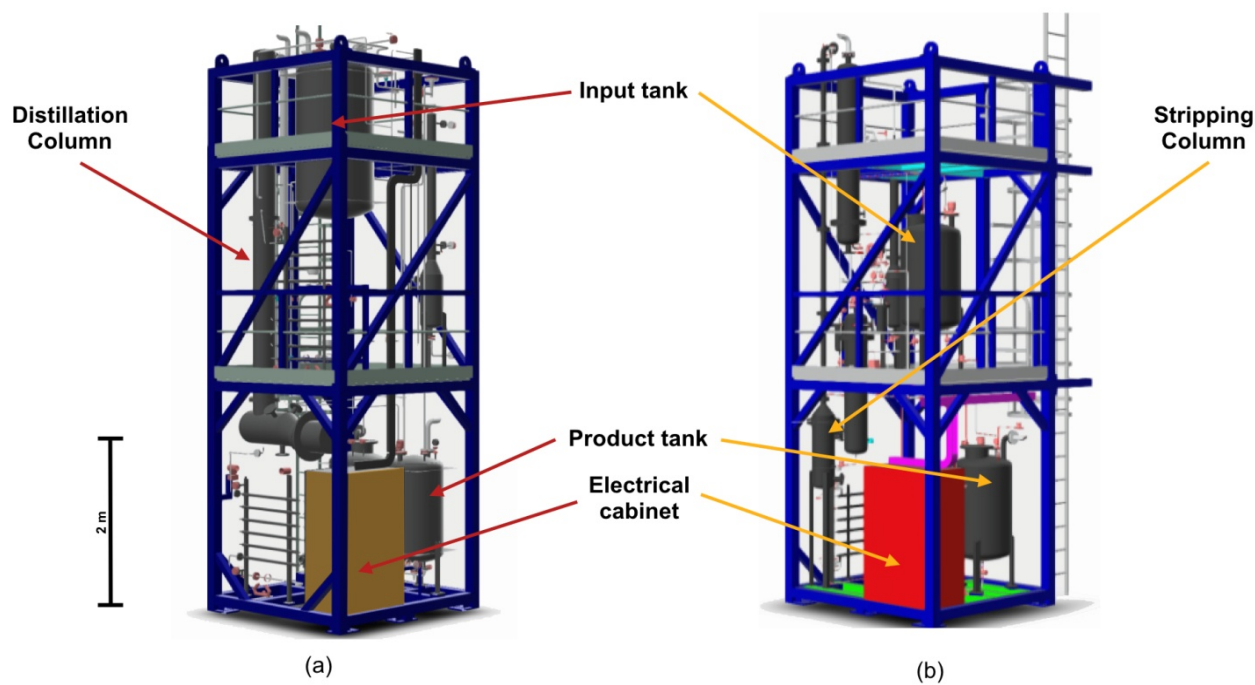

**Fig. 5.** 3D drawing of the distillation plants skid (a) and stripping plant skid (b). The plants are mounted inside a blue skid that can fit a standard ISO container for transportation. They are divided in three floors: in the top floor are mounted the vacuum pumps and the input tanks while in the bottom the product tanks in order to minimize the usage of pumps. The distillation column and the stripping column are placed on a side of the skids and they run from the top floor to the bottom floor to minimize the space required for the installation. In the bottom floor, it is enlightened the electrical cabinet containing the connection for the heaters and pumps power supply and for the CPU of the slow control system receiving the signals from the instruments.

Between February 2016 and March 2016, distillation and stripping pilot plants, under nitrogen atmosphere, were crated in a container and shipped to Shenzhen, China, by sea. One month later, they arrived at the Daya Bay laboratory. After the skids were mounted, all the final connections were made, including the connections to the process lines in Hall 5 of Daya Bay Underground Laboratory.

Before the detector filling each plant has been operated in internal loop mode (described in sec. 2.1 and 2.2) to ensure that they work properly and to adjust the process parameters. During these steps, some problems on the level sensors were identified and solved with a re-calibration of the instruments via HART communicator.

The main features investigated during the commissioning phase were the discharge process of the LAB from the bottom of the distillation column and the thermodynamic parameters that insure a stable and efficient functioning of the stripping column. In particular, regarding the first item it was decided to avoid a continuous discharge of liquid from the bottom of the distillation column because the magnitude of the flow would have been lower than the minimum value measurable by the flow meter.

Regarding the distillation plant, it was decided to further decrease the pressure inside the column in order to reduce the temperature of the LAB and avoid any degradation of the organic compound. In total, around 4000 l of LAB has been distilled and stripped for plant commissioning and final self-cleaning.

After these tests, the plants were connected with Alumina oxide and Water Extraction purification systems through the interconnection system, to the goal of testing the complete purification chain. By reference, Alumina Column plant is based on absorption technique on high quality alumina powder to remove optical impurities and increase the attenuation length of LAB [29] while Water Extraction column is based on the "Scheibel column" design and is intended to remove radioactive contaminants like $^{238}$U, $^{232}$Th and $^{40}$K [29]. These plants are in charge of the Chinese part of the collaboration and they are not described in this paper.

## 5   Results

The performances of the commissioning phase of the distillation and stripping pilot plants are assessed by measuring the remaining content of radio impurities in the LAB and its absorption spectra evaluated after each purification process. The effectiveness of these purification methods in removing the radio impurities cannot be measured by laboratory tests, giving only generic hints on their efficacy. The Daya Bay detector, instead, enables the quantitative evaluation of the residual background in the LAB, which will be reported in the paper describing the full procedure of tests and measurements performed on the whole sets of pilot plants at Daya Bay.

However, meaningful preliminary indications of the effectiveness of the plants can be gathered indirectly through the inspections of the absorption spectra. Indeed, the LAB attenuation length and the absorption spectra were measured before filling the detector and after each purification step [29].

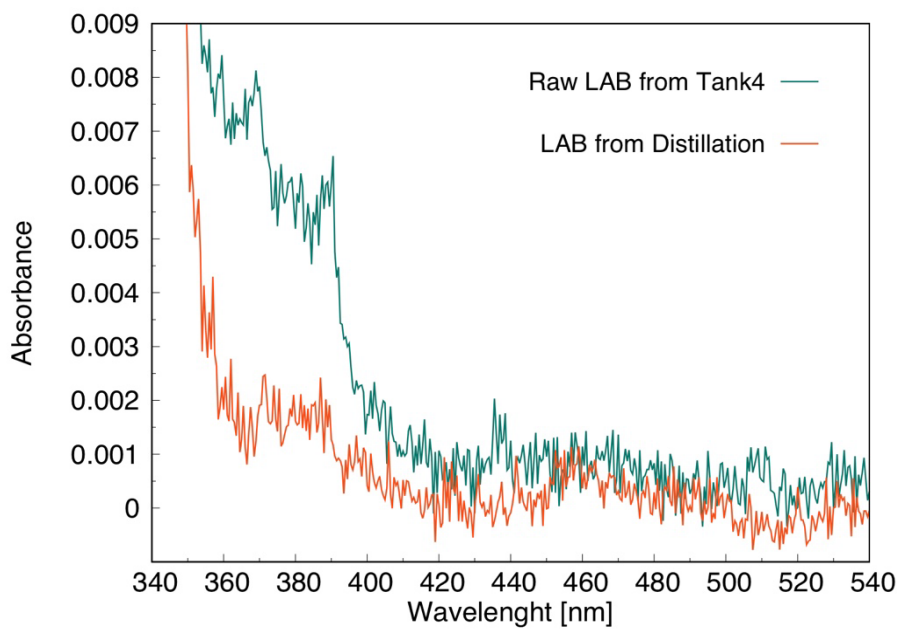

**Fig. 6**. Comparison of the absorption spectra of raw and distilled LAB (modified from [29]). It is important to notice that even if the most reduction of the optical impurities is carried out by the alumina plant, the distillation has a small effect on reducing the attenuation length in the wavelength region around 365 nm.

In Fig. 6, the absorption spectrum is reported as a function of the wavelength (where on abscissa there is the wavelength in nm and on y-axis the absorbance in arbitrary unit). By comparing the spectrum of the raw LAB with the one after distillation, we can infer the very high effectiveness of the distillation plant to remove optical impurities over the whole region of interest.

Moreover, from [29], it is possible to conclude that the stripping procedure, intended to remove gaseous compound and hence not expected to affect the absorption spectrum, is clean enough not to spoil the optical quality as obtained from the previous distillation step.

# 6 Conclusion

This paper described the features and the commissioning phase of a distillation and a stripping pilot plant designed to test the purification efficiency of this processes for a LAB based liquid scintillator in terms of removal of radio and optical impurities. Moreover, the study permitted to evaluate the model built for the calculation of the total reliability of the two pilot plants. For the first time, well-established technologies are integrated for the purification of a LAB based LS. The purification effectiveness, the safety of the plants and of the operators are guaranteed adopting the peculiar features summarized below:

- Using the distillation column input feed (LAB) as a cooling fluid in the total condenser (Fig. 1) leads to a substantial reduction of the energy consumed for the liquefaction of the LAB vapor and for the warm-up of the input feed. Moreover, positioning the condenser (pre-heater) on the top of the column implies a substantial reduction of the plant size.
- The installation inside the distillation column of sieve trays allows to maximize the contact surface between the liquid and vapor phase keeping a high cleanliness level and in turn to get a greater efficiency of the distillation.
- The LAB thermal degradation is reduced by performing the distillation under vacuum with lower boiling temperature.
- Using a variable mixture of steam and nitrogen as gas stream in the stripping column leads to better results on purification efficiency due to the lower $^{222}$Rn content in ultra-pure water, as compared to regular nitrogen. Moreover, since the steam is completely liquefied in the vacuum line condenser and the water disposed properly, a dedicated exhaust system is not necessary.
- While the stripping process has no effect on the optical property of the LAB, the distillation increases the attenuation length in the wavelength region of interest (Fig. 6). The attenuation length measured on scintillator (LAB + 2.5 g/l PPO and 7 mg/l bisMSB) after all the purification process reaches a value of 20 m @ 430 nm, greater than typical values obtained in previous neutrino experiments (Table 3). The attenuation length of pure LAB reaches 25 m @ 430 nm after distillation.
- Adopting the data from the pilot plants, the reliability study for the future JUNO purification plants shows an average of greater than 3 months of MTBF (Table 6). The JUNO distillation plant will be more subject to failure due to its greater complexity and number of components. This model will give also an indication on hierarchy of the most fragile parts of the system that will need a prompt back-up solution in case of failure.

In the perspective of the realization of JUNO, as well as for future massive neutrino experiments, the distillation and stripping processes are expected to play a key role in reducing the radio background contamination and in increasing the attenuation length of the LS.

**Acknowledgments**

This work was partially supported by National Institute for Nuclear Physics (INFN) through the JUNO experiment and by Università degli Studi di Ferrara under the scientific project FIR-2017. The authors would like to thank Mario Masetto, Isabella Canesi and Gabriele Milone and the Polaris staff for their contributions on design and building the plants and Zhou Li, Hu Tao, Yu Boxiang, Cai Xiao, Fang Jian, Lijun Sun and Yuguang Xie from IHEP for their invaluable help during the test operation of the distillation and stripping plants.